\documentclass{article}  
\usepackage{bigsky2007}
\usepackage{graphicx}
\frompage{000} \topage{000}                                              
\def\rightmark{Two-particle correlations with strange baryons and mesons at RHIC}
\def\leftmark{Jana Bielcikova et al.}
 
\title{Two-particle correlations with strange baryons and mesons at RHIC} 
\authors{
{Jana Bielcikova$^1$ for the STAR Collaboration 
}\\[2.812mm]
{\normalsize
\hspace*{-8pt}$^1$ Department of Physics, Yale University, P.O.Box 80124, New Haven, CT-06520, USA \\ 
\\[0.2ex] 
}}
 
\abstract{We present results on two-particle correlations using 
singly-strange particles ($\Lambda$, $\bar{\Lambda}$, $K^0_S$) and 
charged hadrons at intermediate $p_T$ in d+Au and Au+Au collisions at RHIC. 
We discuss properties of the near-side correlation peak in both azimuthal 
and pseudo-rapidity space and separate jet-like contributions 
from the long-range pseudo-rapidity correlations (the ridge). In particular, 
we study the centrality and $p_T$ dependence of the jet 
and ridge yields for various trigger and associated particle 
species and compare the results to model predictions.}

 
\begin{document}

\maketitle

\section{Introduction}
Heavy-ion collisions at relativistic energies provide a unique tool for studies 
of strongly interacting matter under extreme conditions 
of high temperature and energy density. Studies of particle production 
at the top RHIC energy ($\sqrt{s_{NN}}$~=~200~GeV) revealed a strong suppression 
of inclusive transverse momentum ($p_T$) distributions of identified 
light hadrons in central Au+Au collisions with respect to p+p, 
d+Au and peripheral Au+Au collisions~\cite{Adler:2003kg,Adams:2003am}. This suppression, commonly
referred to as jet quenching,  reaches in central Au+Au collisions a value 
of about 0.2 and is present out to large transverse momenta ($p_T\approx$~12~GeV/$c$). 
This observation is accompanied by enhanced baryon production~\cite{Adams:2006wk} 
in the intermediate-$p_T$ range of $\approx$2-6~GeV/$c$. 
The measured baryon/meson ratios in both non-strange and strange quark sectors 
increase with $p_T$ up to about $p_T\approx$~3~GeV/$c$, where the enhancement of 
baryon/meson production reaches its maximum value of $\approx$~3 relative to  p+p collisions. 
A fall-off of the baryon/meson ratio is observed for $p_T>$~3~GeV/$c$ and
both, non-strange and strange baryon/meson ratios, approach each other 
and eventually reach the values measured in p+p collisions at $p_T\approx$~6~GeV/$c$.  

These findings suggest that a dominant source of particle production at intermediate $p_T$ 
is not from jet fragmentation, rather, parton recombination and coalescence models have been suggested 
as alternative mechanisms~\cite{Fries:2003kq,Greco:2003mm,Greco:2003xt,Hwa:2002tu}.
In these models, competition between recombination and fragmentation results in a shift 
of the onset of the perturbative regime to higher transverse momenta $p_T\approx$~4-6~GeV/$c$.
Due to the steeply falling parton transverse momentum spectrum, the fragmentation 
process is a much less efficient particle production mechanism than recombination. 
Moreover, the steeply falling parton spectrum favors recombination of three quarks to 
form a baryon over the recombination of two quarks to form a meson with the same $p_T$. 
Such mechanisms then naturally lead to an increased production of baryons 
relative to mesons which is in qualitative agreement with the data.

In addition to the inclusive measurements, studies of two-particle correlations 
of charged hadrons in Au+Au collisions revealed the presence of additional 
long-range pseudo-rapidity correlations on the near-side commonly 
referred to as  the {\it ridge}~\cite{PutschkeHPQM06}. Such extended 
pseudo-rapidity correlations are absent in p+p and d+Au collisions.

It is expected that studies of two-particle azimuthal correlations using identified particles 
will help to understand processes relevant for particle production at intermediate $p_T$ and give insight into the origin of the ridge. The wealth of data collected by the STAR experiment 
at $\sqrt{s_{NN}}$~=~200~GeV 
offers the possibility of a detailed study of strangeness production up to high $p_T$.
In this paper, we discuss the properties of two-particle correlations 
using strange particles ($\Lambda$, $\bar\Lambda$, $K^0_S$) as well as unidentified charged particles 
in d+Au and Au+Au collisions.  We focus our discussion on 
properties of the near-side correlation peak in both azimuthal 
and pseudo-rapidity space and separate jet-like contributions 
from the long-range pseudo-rapidity correlations. In particular, 
we study the centrality and $p_T$ dependence of the jet 
and ridge yields for various trigger and associated particle 
species and compare the results to model predictions.

\section{STAR experiment}
STAR is a large acceptance, multi-purpose spectrometer consisting
of several detectors inside a large solenoidal magnet with a magnetic 
field of 0.5~T. The analysis presented in this paper is based exclusively 
on charged particle tracks detected and reconstructed in the Time Projection Chamber 
(TPC). The TPC is well suited for the correlation studies 
because of its full azimuthal coverage. The TPC provides 
up to 45 independent spatial and energy loss ($dE/dx$) measurements along each charged particle track. 
The momentum resolution is determined to be $\Delta k/k\sim0.0078+0.0098\cdot p_T$~(GeV/$c$), where $k$ is 
the track curvarture proportional to $1/p_T$. 
More details can be found elsewhere~\cite{Anderson:2003ur}.

  The weakly decaying strange particles (V0s) are reconstructed by a topological 
analysis from their decay products measured in the TPC. The geometrical cuts have 
been optimized to achieve a very clean sample of the V0 particles. 
Above $p_T>$~2~GeV/$c$ the signal to background ratio with these cuts is approximately 15:1.

 The results presented in this paper are based on the d+Au data set taken in 2003 and Au+Au data set 
measured in 2004 at $\sqrt{s_{NN}}$~=~200~GeV.

\section{Data analysis}

In the following paragraphs, we discuss properties of the near-side azimuthal
correlations using identified strange particles ($\Lambda$, $\bar{\Lambda}$, and K$^0_S$)
and  unidentified charged particles either as a trigger or an associated particle. 
The azimuthal distributions among trigger and associated particles in a given centrality bin 
are defined as: 
\begin{equation}
\label{corr}
 C(\Delta\phi)=\frac{1}{N_{trigger}}\int\!\!\!\int\!\!\!\int\!\! dp_T^{trigger}dp_T^{associated}d(\Delta\eta) \frac{dN_{pair}(p_T^{trigger},p_T^{associated},\Delta\phi,\Delta\eta)}{\epsilon(p_T^{associated})}, 
\end{equation}
where $\Delta\phi=\phi^{trigger}-\phi^{associated}$, $N_{trigger}$ is the number of trigger particles, 
and $\epsilon$ is the reconstruction efficiency of associated particles. The correlation functions 
are also corrected for the triangular acceptance in $\Delta\eta$ and TPC sector boundaries in azimuth.

The data are fit with two Gaussians on top of a flat background in d+Au 
and elliptic flow ($v_2$) modulated background in Au+Au collisions, respectively. 
The yield of associated particles is calculated as the area under the Gaussian peak.
We study separately the jet and ridge contributions to the near-side yield by analyzing the correlations in two 
$\Delta\eta$ windows: $|\Delta\eta|<$~0.7 containing both jet and ridge correlations, and $|\Delta\eta|>$~0.7 
containing only the ridge contributions, assuming the jet contribution at large $\Delta\eta$ is negligible. 
Assuming  uniformity of $v_2$ with $\eta$, the jet yield is free of systematic uncertainties due to the elliptic 
flow subtraction. For the ridge yield, these systematic errors are estimated  by subtracting 
the $v_2$ measured by the event plane method (the lower bound) and by the 4-particle cumulant method (the upper bound).
The uncertainties in the elliptic flow subtraction result in about a 30$\%$ 
systematic error on the extracted associated yield. 

\begin{figure}[t!]
\begin{center}
\begin{tabular}{lr}
\hspace{-1.5cm}
\includegraphics[height=6.0cm]{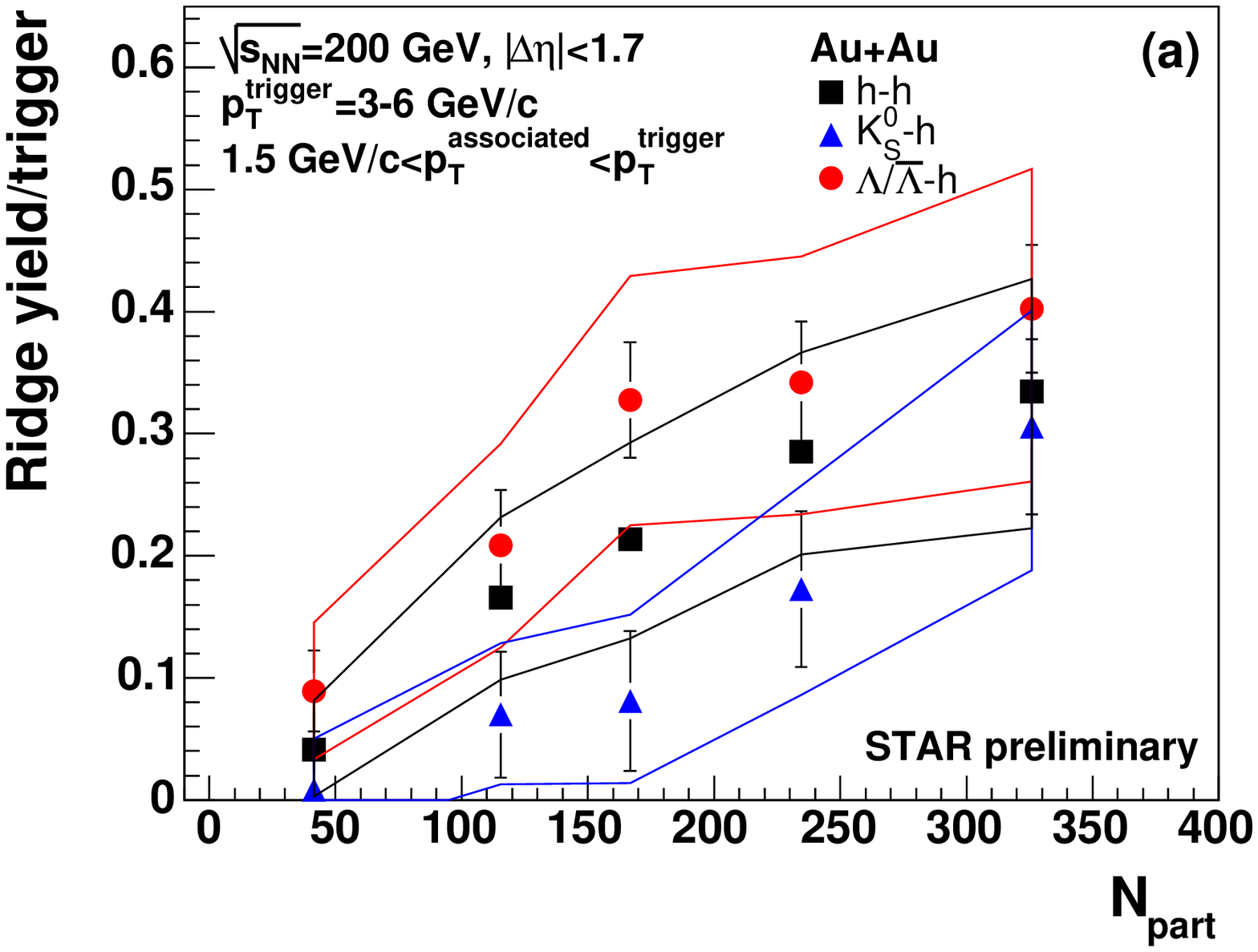}
&
\hspace{-0.8cm}
\includegraphics[height=6.0cm]{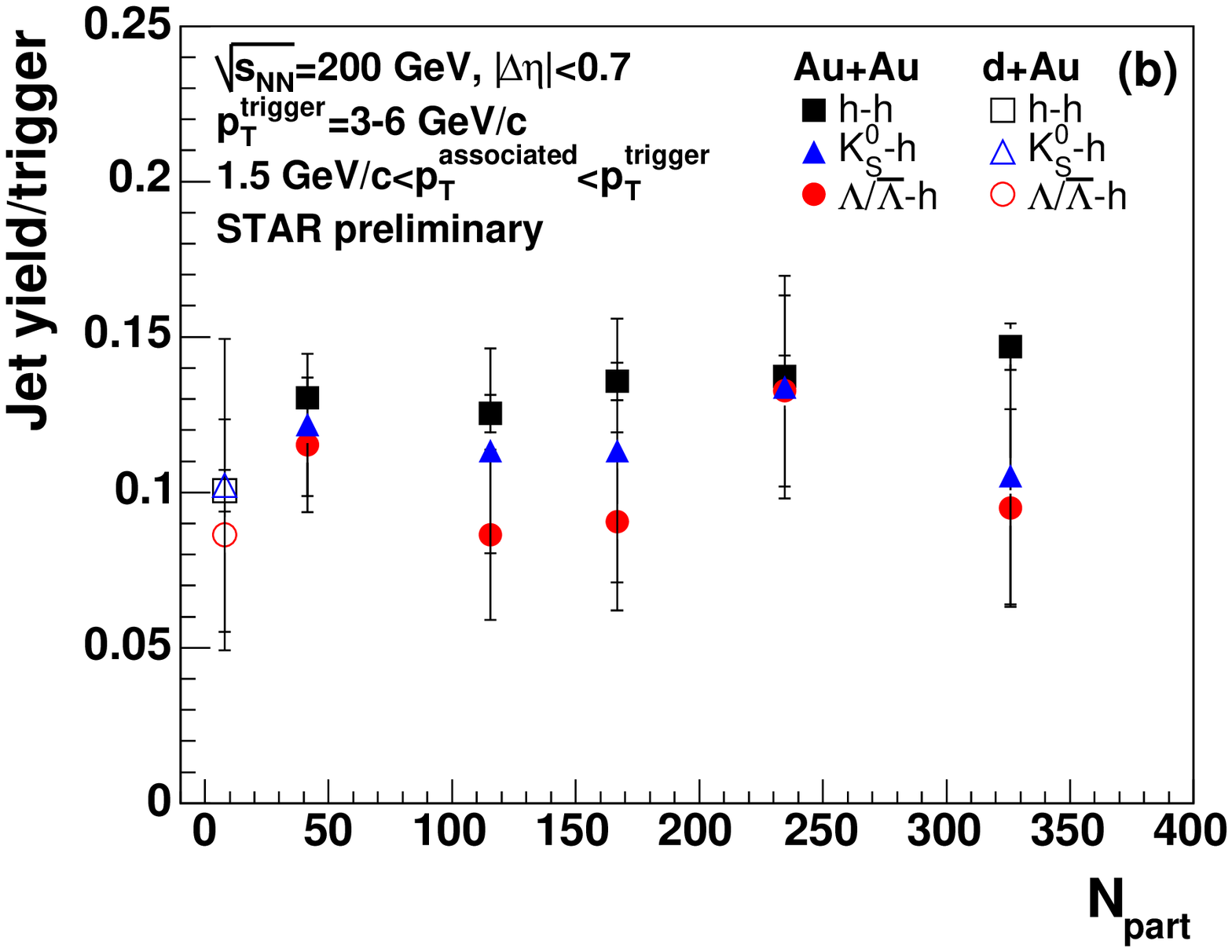}
\end{tabular}
\end{center}
\caption{Centrality dependence of the ridge yield (a) and jet yield (b) of associated charged particles for various trigger species in d+Au and Au+Au collisions. The error bands indicate systematic errors on the ridge yield due to the $v_2$ subtraction. } 
\label{rjcentr}
\end{figure}

\begin{figure}[h!]
\begin{center}
\begin{tabular}{lr}
\hspace{-1.5cm}
\includegraphics[height=6.0cm]{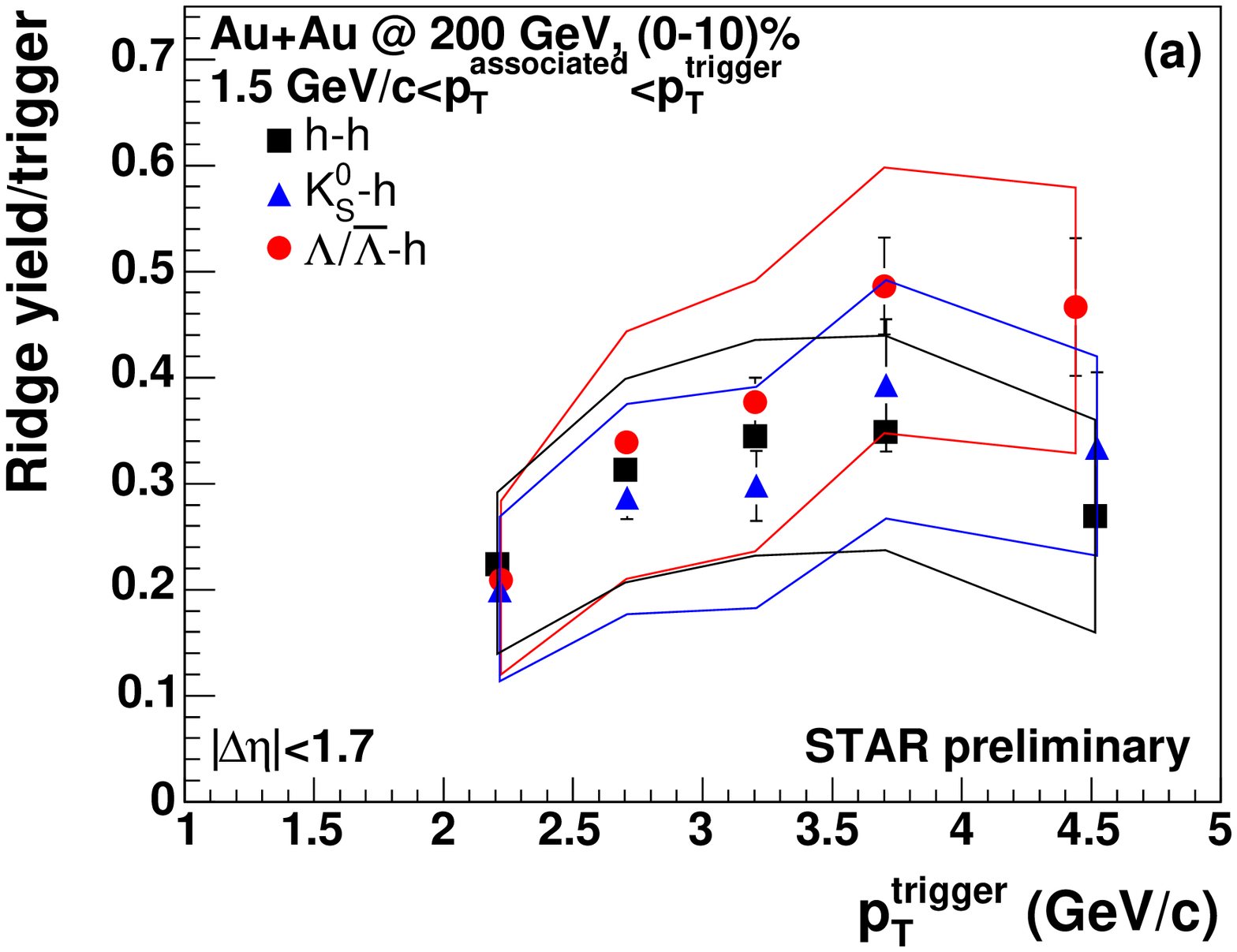}
&
\hspace{-0.8cm}
\includegraphics[height=6.0cm]{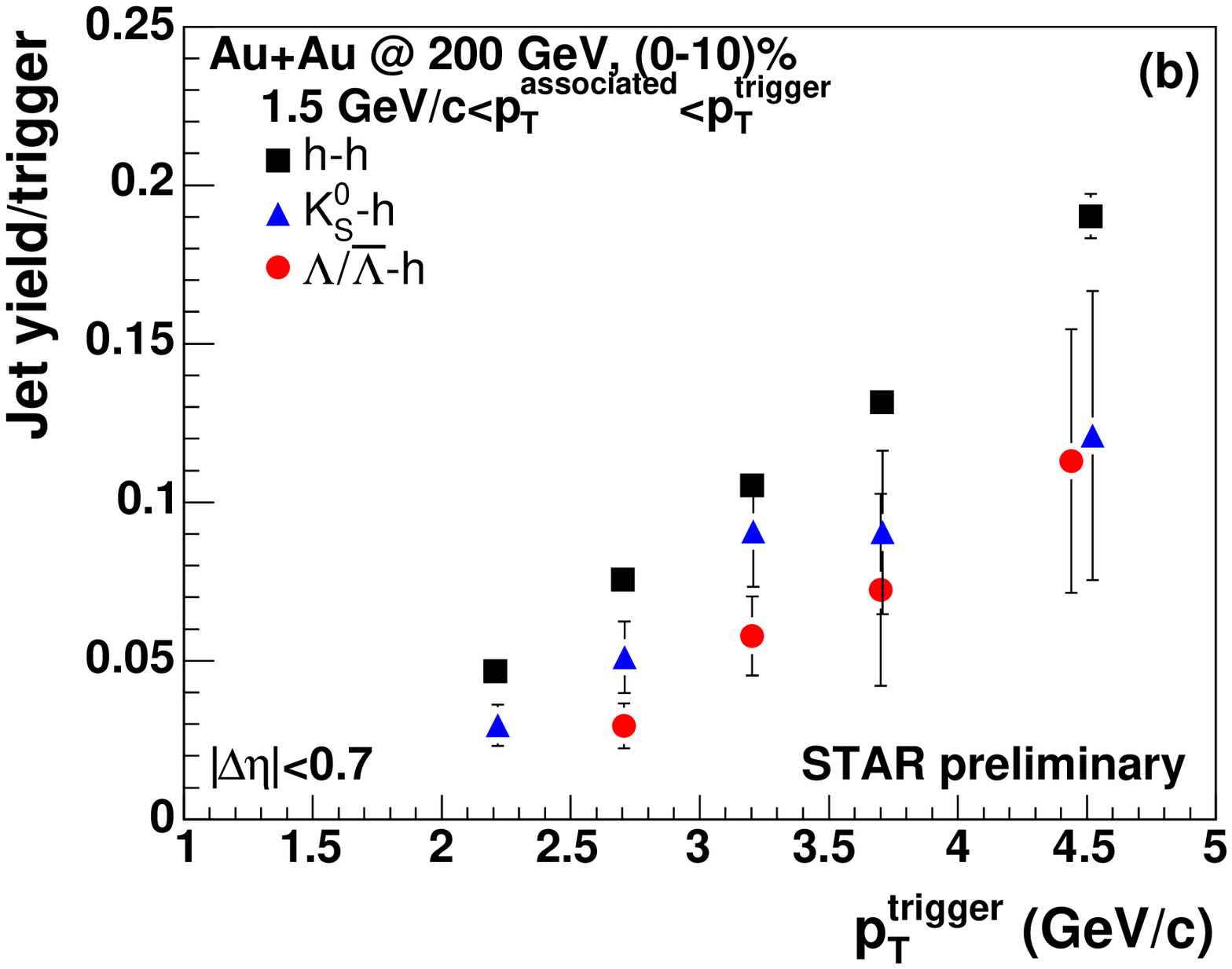}
\end{tabular}
\end{center}
\caption{Dependence of the ridge yield (a) and jet yield (b) on $p_T^{trigger}$ for various trigger species in central (0-10\%) Au+Au collisions. The bands indicate systematic errors on the ridge yield due to the $v_2$ subtraction.}
\label{rjpttrig}
\end{figure}

\section{Results}
We first discuss properties of correlations using identified trigger particles which are 
associated with charged particles. Comparing d+Au and Au+Au collisions, we observe 
an increase of the near-side yields by a factor of 3-4 going from d+Au to central 
Au+Au collisions for all studied trigger species.  Figure~\ref{rjcentr} 
shows separately the centrality dependence of the ridge and jet yields of the near-side correlation. 
While the jet yield is approximately independent of centrality and consistent with its value in d+Au, 
the ridge yield increases steeply with centrality. 

\begin{figure}[b!]
\begin{tabular}{lr}
\includegraphics[height=8.5cm]{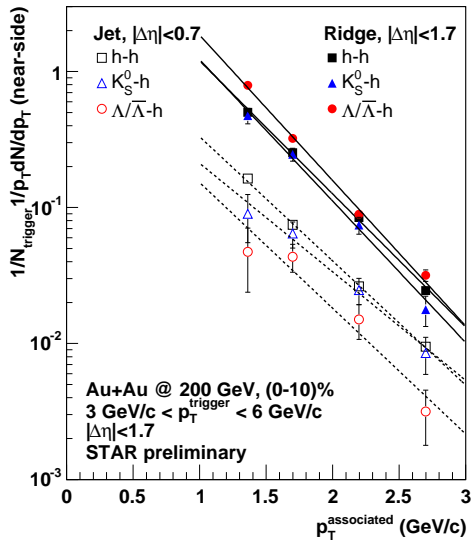}
&
\begin{minipage}[t]{0.5\linewidth}
\vspace{-8.0cm}
\caption{Invariant $p_T$ distribution of associated charged particles in the jet and ridge 
in central Au+Au collisions for various trigger species indicated by the legend. 
The curves are exponential fits to the data.}
\label{rjptassoc}
\end{minipage}
\end{tabular}
\end{figure}
 The observed increase of the near-side yield with centrality is qualitatively 
in line with the recombination model expectations and points toward 
a significant role of thermal-shower recombination in Au+Au  collisions~\cite{Hwa:2005ui}.
Because the long range pseudo-rapidity correlations play a significant role 
in Au+Au collisions in the studied $p_T$ range, a check of how well the individual 
jet and ridge contributions are reproduced in the model is required as well. More detailed discussion
on possible mechanisms of ridge origin is given in Section~\ref{sumsec}.

\begin{table}[t!]
\begin{center}
\begin{tabular}{|c|l|l|} 
\hline\\[-10pt]
Trigger particle		& T(ridge) MeV	 	& T (jet) MeV \\ \hline
h$^{\pm}$	 		& 438$\pm$4 (stat.) 	& 478$\pm$8   \\ \hline
$K^0_S$	 			& 406$\pm$20 (stat.)	& 530$\pm$61    \\ \hline
$\Lambda$ 			& 416$\pm$11 (stat.)	& 445$\pm$49    \\ \hline
\end{tabular}
\caption{Inverse slope of associated particle spectra for jet and ridge in central Au+Au collisions.}
\label{tab-inverse-slope}
\vspace*{-14pt}
\end{center}
\end{table}

\begin{figure}[b!]
\begin{center}
\includegraphics[height=7.5cm]{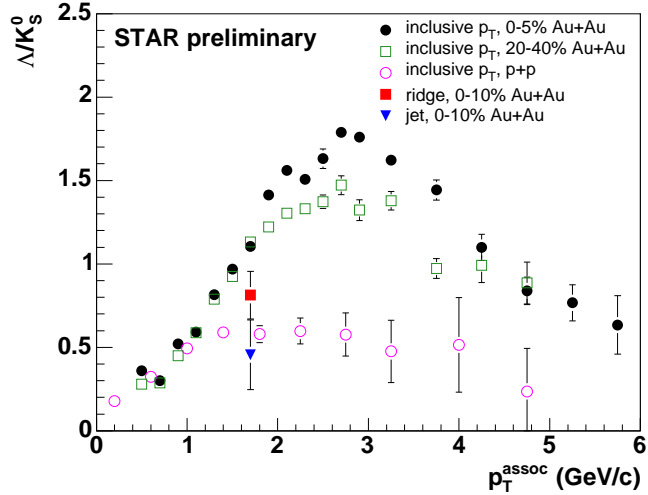}
\caption{$\Lambda$/$K^0_S$ ratio measured in inclusive $p_T$ distributions, near-side jet and ridge-like 
correlation peaks in Au+Au collisions together with this ratio obtained from inclusive $p_T$ spectra in p+p collisions.}
\end{center}
\label{bmratio}
\end{figure}

Next, we study the dependence of the jet and ridge yields on the transverse momentum of the trigger particle,
$p_T^{trigger}$, shown in Figure~\ref{rjpttrig}. While the ridge yield increases with $p_T^{trigger}$ 
and flattens off for $p_T^{trigger}>$~3~GeV/$c$, the jet yield
keeps increasing with $p_T^{trigger}$ as expected for jet-like processes. 
The jet yield for $\Lambda$ triggers is systematically below that of charged hadron and $K^0_S$ triggers. 
This can be partially attributed to the larger width of the baryon triggered correlations with respect 
to the meson triggered case. In addition, remaining effects of artificial track merging, which result 
in the yield loss, affect V0s more than charged tracks. These effects are currently under investigation.

We have also measured the invariant $p_T$ spectra of associated charged particles 
for $p_T^{trigger}$~=~3-6~GeV/$c$ in central Au+Au collisions as shown in Figure~\ref{rjptassoc}.
We have fit the distributions with an exponential function $e^{-p_T/T}$ and extracted the inverse 
slope, $T$, given in Table~\ref{tab-inverse-slope}.
The ridge spectra have, averaged over all studied trigger species, $T$~=~420$\pm$8~MeV, 
close to that of the bulk when the fit is performed in the same $p_T$ range. 
The jet spectra have $T$~=~480$\pm$30~MeV. A detailed investigation 
of the evolution of the inverse slope of the jet and ridge spectra with $p_T^{trigger}$ has been carried out 
 with charged trigger and associated particles~\cite{PutschkeHPQM06}, where the statistics are richer
than for identified correlations. A clear difference in the evolution of the slope with $p_T^{trigger}$
was observed for jet and ridge-like correlations. While the inverse slope of the ridge spectra remains 
constant with increasing $p_T^{trigger}$ and differs from the bulk by about ~50~MeV, the jet spectra show 
a steep increase of the slope with $p_T^{trigger}$.    

It is interesting to study the composition of particles in the ridge and jet in order to look 
for a possible enhancement of the baryon/meson ratio which is present in the inclusive 
$p_T$ distributions in Au+Au collisions. We have carried out a preliminary study 
of two-particle correlations using charged trigger particles with 
$p_T^{trigger}$~=~2-3~GeV/$c$ which were associated with identified 
strange particles with 1.5~GeV/$c$~$<p_T^{associated}<p_T^{trigger}$. 
Using the same method as above, we have extracted the jet 
and ridge yields and calculated the $\Lambda$/$K^0_S$ ratio.
The obtained results on this baryon/meson ratio for particles associated with the jet 
and ridge are shown in Figure~\ref{bmratio} together with the ratio 
measured from the inclusive $p_T$ spectra. 
The $\Lambda$/$K^0_S$ ratio calculated for the jet 
is 0.46$\pm$0.21 and consistent with that measured in p+p. 
The same ratio in the ridge is 0.81$\pm$0.14, higher than in the jet. 
The large statistical errors do not allow to draw a definite conclusion on the origin 
of the ridge yet. 

\section{Summary and discussion}
\label{sumsec}
We have reported results on the properties of the near-side correlation peak 
at intermediate-$p_T$ at RHIC using strange particles and unidentified charged hadrons. 
The measured correlations reveal a strong contribution from the long-range $\Delta\eta$ 
correlations at near-side for all studied particle species. The strength of these 
ridge-like correlations increases steeply with centrality and the inverse slope of the associated particle 
spectra is, by $\approx$~50~MeV, higher than that of particles produced in the bulk. More data and studies
are needed to draw final conclusions on the baryon/meson ratios in the ridge.
The jet yield is, contrary to this, independent within errors of centrality in Au+Au 
collisions and consistent with that in d+Au collisions. In addition, the baryon/meson composition 
in the jet in Au+Au is close to that measured in p+p collisions.

While the existence and properties of the jet-like correlations are as those
in p+p and d+Au collisions, the origin of the ridge is not fully understood. This phenomenon  
has triggered significant interest in the theoretical community and several models 
are currently available to describe the observed ridge. Below we discuss their main
features.
 
A model based on the recombination and fragmentation of partons in the medium \cite{Chiu:2005ad}, 
links the origin of the ridge to the longitudinal expansion of the thermal partons that are enhanced 
by the energy loss of a hard parton traversing the medium.  The model predicts that the baryon/meson 
ratio of particles associated with the ridge should follow that of particles produced in the bulk and 
therefore lead to an increased production of baryons. 
In addition, the model predicts the inverse slope of the enhanced thermal 
parton distribution to be only by 15~MeV larger than that of particles in the bulk.  The data show 
that over a broad range of $p_T^{trigger}$~=~3-12~GeV/$c$, this difference is $\approx$50~MeV.

Another suggested mechanism for the ridge origin is based on jet quenching and strong radial flow~\cite{Voloshin:2004th}. The radial expansion of the system is predicted to create strong position-momentum correlations that lead to characteristic rapidity, azimuthal and $p_T$ correlations among produced particles. 
 
In another approach, the interaction of high-$p_T$ partons with a dense medium 
under the presence of strong longitudinal collective flow is predicted to lead to a characteristic 
breaking of the rotational symmetry of the average jet energy and multiplicity distribution 
in the $\eta\times\phi$ plane~\cite{Armesto:2004pt}. This will in turn cause a medium-induced broadening 
of gluon radiation in pseudo-rapidity and form a ridge in $\Delta\eta$.

The effects of momentum broadening of a heavy quark in an anisotropic plasma 
have also been studied~\cite{Romatschke:2006bb}. It is shown that the momentum broadening, induced by collisional 
energy loss in the leading logarithmic approximation, is more pronounced along the longitudinal 
direction that in the reaction plane. Unfortunately, in its current form, this model is not directly 
applicable to our data. First of all, the ridge has been observed for light hadrons while the calculation is done for heavy quarks only. Secondly, the large value of the momentum space 
anisotropy parameter required by the calculation to describe the data is incompatible with the low shear viscosity of the medium at RHIC as pointed out elsewhere~\cite{Majumder:2006wi}.

The mechanism proposed in~\cite{Majumder:2006wi}, relates the origin of the ridge to the longitudinal expansion of the medium and spontaneous formation of extended color fields in such expanding medium due to the presence of plasma instabilities. The predicted ridge at low transverse momenta is expected to decrease with increasing energy of the associated radiation. The momentum range of the partons contained in the ridge is in the recombination regime and therefore the authors of this model predict that it would reflect itself in the baryon to meson ratio of associated hadrons.

In order to draw final conclusions about the origin of the ridge, more quantitative theoretical 
calculations are needed.

\bibliography{bigsky2007-template}

\begin{thebibliography}{10}

\bibitem{Adler:2003kg} S. S. Adler {\it et al} (PHENIX), {\it Phys. Rev. Lett.} {\bf 91} (2003) 172301.
\bibitem{Adams:2003am} J. Adams {\it et al} (STAR), {\it Phys. Rev. Lett.} {\bf 92} (2004) 052302.
\bibitem{Adams:2006wk} J. Adams  {\it et al} (STAR),  nucl-ex/0601042.
\bibitem{Abelev:2006jr} B. Abelev {\it et al.} (STAR), {\it Phys. Rev. Lett.} {\bf 97} (2006) 152301.
\bibitem{Fries:2003kq} R. J. Fries, B. Mueller, C. Nonaka, and S. A. Bass {\it Phys. Rev.} {\bf C68} (2003) 044902.
\bibitem{Greco:2003mm} V. Greco, C. M. Ko  and P. Levai {\it Phys. Rev.} {\bf C68} (2003) 034904.
\bibitem{Greco:2003xt} V. Greco, C. M. Ko  and P. Levai, {\it Phys. Rev. Lett.} {\bf 90} (2003) 202302.
\bibitem{Hwa:2002tu} R. C. Hwa and C. B. Yang {\it Phys. Rev.} {\bf C67} (2003) 034902.
\bibitem{PutschkeHPQM06} J. Putschke (STAR), {\it Nucl. Phys.} {\bf A783} (2007) 507; nucl-ex/0701074.
\bibitem{Anderson:2003ur} M. Anderson (STAR), {\it Nucl. Instrum. Meth.} {\bf A499} 659.
\bibitem{Hwa:2005ui} R. C. Hwa and Z. Tan, {\it Phys. Rev.} {\bf C72} (2005) 057902.
\bibitem{Chiu:2005ad} C. B. Chiu, R. C. Hwa, {\it Phys. Rev. C} {\bf 72} (2005) 034903.
\bibitem{Voloshin:2004th} S. A. Voloshin, {\it Nucl. Phys.} {\bf A749} (2005) 287; nucl-th/0312065.
\bibitem{Armesto:2004pt} N. Armesto, C. A. Salgado, U. A. Wiedemann, {\it Phys. Rev. Lett.} {\bf 93} (2004) 242301.
\bibitem{Romatschke:2006bb} P. Romatschke, {\it Phys.Rev.} {\bf C75} (2007) 014901.
\bibitem{Majumder:2006wi} A. Majumder, B. Mueller, S. A. Bass, hep-ph/0611135.
   
\end{thebibliography}
\bibliographystyle{bigsky2007}
 

\end{document}